\documentclass[aps,twocolumn,prl,color,psfig,epsf,sort&compress,floatfix]{revtex4}

\usepackage{amsmath}
\usepackage{graphics}
\usepackage[T1]{fontenc}
\usepackage[utf8]{inputenc}
\usepackage{color}
\usepackage{graphicx}

\begin{document}
\title{Soft channel formation and symmetry breaking in exotic active emulsions}

\author{L.N. Carenza $^1$, G. Gonnella$^1$, A. Lamura $^4$, D. Marenduzzo $^2$, G. Negro $^{*2}$, A. Tiribocchi $^3$ }
\affiliation{$^1$Dipartimento di Fisica, Università degli Srudi di Bari and INFN, Sezione di Bari, Via Amendola 173, 70126 Bari, Italy, $^2$ SUPA, School of Physics and Astronomy, University of Edinburgh, Edinburgh EH9 3JZ, United Kingdom, $^3$ Center for Life Nano Science@La Sapienza, Istituto Italiano di Tecnologia, 00161 Rome, Italy, and IAC - CNR, via dei Taurini 19, Rome, Italy, $^4$IAC - CNR, Via Amendola, 122/D 70126 Bari - Italy,$^*$giuseppe.settimio.negro@gmail.com}

\begin{abstract}
We use computer simulations to study the morphology and rheological properties of a bidimensional emulsion resulting from a mixture of a passive isotropic fluid and an active contractile polar gel, in the presence of a surfactant that favours the emulsification of the two phases. By varying the intensity of the contractile activity and of an externally imposed shear flow, we find three possible morphologies. 
For low shear rates, a simple lamellar state is obtained. For intermediate activity and shear rate, an asymmetric state emerges, which is characterized by shear and concentration banding at the polar/isotropic interface. A further increment in the active forcing leads to the self-assembly of a soft channel where an isotropic fluid flows between two layers of active material. 
We characterize the stability of this state by performing a dynamical test varying the intensity of the active forcing and shear rate. Finally, we address the rheological properties of the system by measuring the effective shear viscosity, finding that this increases as active forcing is increased -- so that the fluid thickens with activity.
\end{abstract}

\flushbottom
\maketitle

\thispagestyle{empty}

\section*{Introduction}

Active matter~\cite{kruse2004,joanny2009,marchetti2013,doostmohammadi2019} has recently become a topical area of research in soft matter physics. Much effort
has been spent to understand the underlying non equilibrium mechanisms that rule 
the dynamics of systems where constituents are actively capable of converting internal 
energy into work.  
Research in this field has been steered with the intention to 
uncover the physics behind the behaviours of self-driven and living systems with 
the important result that a vast range of phenomena have been discovered in actual
experiments~\cite{Sokolov2009,DiLeonardo2010,Yeomans2014,doostmohammadi2018,needleman2017} and replicated by means of numerical simulations at a high pace.
This trend has been fostered by many unexpected findings, ranging from 
spontaneous flow~\cite{simha2002,kruse2004,marenduzzo2007,PhysRevLett.101.198101} and active turbulence~\cite{dombrowski2004,wensink2012,carenza19turb,PhysRevLett.122.214503,thampi2016active} to motility-induced phase separation~\cite{cates2015, gonnella2015, digregorio2018}.
Despite a comprehensive theory is still missing, this exploratory era 
is gradually giving way to more applicative research aimed at
gaining control of active matter to exploit its properties for \emph{in vivo} drug-delivery and in technological applications, such as novel smart materials.


An important candidate for the development of activity-based micro-devices is represented by active liquid crystals~\cite{doostmohammadi2018}. These may be either of biological origin (high-density bacterial suspensions  and cytoskeletal extracts)~\cite{dombrowski2004,zhang2010,surrey2001,prost2015,guillamat2016,sanchez2012} or synthetic (Janus particles or polyacrylic acid hydrogels (PAH)). 
The active constituents are elongated and/or filamentous and exhibit the tendency to mutually align to form long-ranged ordered domains. 
This feature, combined with small scales energy injection, is
able to excite self-sustained flows in the underlying fluid where the active units are suspended --a regime commonly addressed as \emph{spontaneous flow}-- or even drive the system towards a chaotic state, named \emph{active turbulence}, in analogy with the well known phenomenon developing in isotropic fluids at high Reynolds numbers. 


Recent experiments and simulations have shown that active fluids can develop interesting rheological behaviours when subject to an externally imposed shear. 
In particular, Lopez \emph{et al.}~\cite{lopez2015} observed that suspensions of \emph{E. choli} bacteria are able to develop a \emph{superfluidic} state characterized by null apparent viscosity, while Guo \emph{et al.}~\cite{guo2018} found transient negative viscosity states.
Active fluids react differently to an externally imposed stress in accordance to the swimming mechanism of the individual constituents which can be broadly classified either as \emph{extensile} (bacteria equipped with flagella as \emph{E. choli}  and microtubule bundles) or \emph{contractile} (acto-myosin suspensions).
Extensile swimmers (or \emph{pushers}) act on the surrounding environment 
as out-warding force dipoles, thus generating  flows in the far field with fluid expelled along the long axis of the swimmer and drawn across its body.
Oppositely, contractile swimmers (or \emph{pullers}) behave as in-warding force dipoles and the flow structure is reversed.
Analytical and numerical works confirmed that extensile suspensions can
strengthen the imposed flow, thus leading to the reduction of the viscosity of the 
fluid, while contractile systems have the opposite 
effect and they increase the viscosity, a phenomenon that is commonly addressed as 
\emph{active thickening}.


Most experiments concerning the rheology of active fluids focused on the
behaviour of a single fluid system, where the active constituents are uniformly distributed. Nevertheless, many experiments have shown that is now possible to encapsulate active material in a water-in-oil emulsion~\cite{sanchez2012, guillamat2016} to have droplets or shells
of active material. Confining active liquid crystals is key to control active fluids and to localize the typical scales of energy injection. Moreover, it constitutes by itself a first step towards the development of micro-cargos for drug-delivery. 


Simulations based on active gel theory have been proved to be able 
to qualitatively replicate the dynamics of active droplets --from the regular motion of topological defects~\cite{doostmohammadi2018,Giomi20130365}, to the turbulent regime~\cite{PhysRevX.5.031003,carenza19turb}-- and predict new motility modes~\cite{Giomi20130365,tjhung2012,tjhung2015,Carenza22065}. In this paper we consider a theoretical model for an active binary mixture,
to numerically study the rheological response of contractile systems under an externally imposed shear. The emulsion is prepared by mixing an isotropic passive fluid with a polar active one in presence of a surfactant which favours the emulsification of the two phases. 
We varied systematically the intensity of the active forcing and the shear rate, integrating the equations by means of a hybrid lattice Boltzmann approach~\cite{denniston2001,negro2018,reviewepje}.
\textcolor{black}{We found that  competition between contractile activity and the imposed shear leads to interesting phenomena:} in particular weak contractile activity is able to order the lamellar pattern when shear rate is absent or highly suppressed, while for more intense active forcing the system undergoes symmetry-breaking in the direction normal to the walls leading to pronounced concentration banding --a phenomenon that we also observed in the passive counterpart at high shear rates. Interestingly, if both external and active forcing are strong enough, the symmetry is restored and a channel of passive isotropic fluid is formed in the middle of the system, while two soft active layers adhere to the wall. We tested the stability of such configuration by varying both activity and shear rate over time, finding that contractile forcing is able to stabilize the channel, even starting from an asymmetric state. Finally, we found that the effective viscosity of the system increases together with the intensity of the contractile activity. 

The article is organized as follows. In the next Section we will present the model used to describe our system (further details on the numerical methods and simulation parameters can be find in the Appendix). Later on, we will provide a morphological characterization of the various states obtained at varying activity and shear rate and we will thoroughly address the dynamics of the symmetry breaking and the soft channel formation. Finally we will present the results of the dynamical study performed varying activity and shear rate over time. The rheological measurements are outlined in the final discussion.

\section*{The model}
In this Section we will present the physical variables relevant for the characterization of the nemato-hydrodynamic state of the system, its equilibrium properties in absence of activity and the evolution equations ruling the dynamics. We introduce the scalar field $\phi({\bf r},t)$ as the concentration of active material and the polarization ${\bf P}({\bf r},t)$ --a vector field accounting for the mean orientation of the active constituents. The velocity of the fluid will be denoted by ${\bf v}({\bf r},t)$ and the density by $\rho({\bf r},t)$.

The equilibrium properties of the system are encoded in a modified Landau-Brazovskii~\cite{gonnella1998,braz,negro2018} free-energy functional $\mathcal{F}[\phi,\mathbf{P}]$ that can be expressed as
\begin{equation}
\label{freeE}
\mathcal{F}[\phi,\mathbf{P}] = \mathcal{F}^{bm}[\phi] + \mathcal{F}^{pol}[\phi,\mathbf{P}] + \mathcal{F}^{anch}[\phi,\mathbf{P}].
\end{equation}
the first contribution defines the thermodynamic property of the binary mixture:
\begin{equation}
\begin{split}
\label{freeE_phi}
\mathcal{F}^{bm}[\phi] = \int d^{2}r\, \biggl\{ & \frac{a}{4\phi_{cr}^4}\phi^{2}(\phi-\phi_0)^2 +\\ &\frac{k_\phi}{2}\left|\nabla \phi\right|^{2}+\frac{c}{2} (\nabla^2\phi)^2 \biggl\}
\end{split}
\end{equation}
The first term allows for the segregation of the two phases when $a>0$, as the free energy has two minima at $\phi=0$ and $\phi_0$, respectively corresponding to the passive and active phase, while $\phi_{cr}=\phi_0/2$. The gradient terms determine the surface tension. Here we choose $k_\phi<0$ so to favour the formation of interfaces throughout the system, while $c$ must be positive to guarantee thermodynamic stability. 
This is equivalent to disperse a suitable amount of surfactant with a relaxation time much smaller than the one of the other two phases.  For a symmetric composition of the mixture the theory defined by $\mathcal{F}^{bm}[\phi]$ admits as ground state a lamellar modulation with wave-number $\kappa = \sqrt{|k_\phi|/2c}$ when $a<k^2_\phi/4c$.
The second contribution to $\mathcal{F}$ is borrowed from the theory for liquid crystals and adapted for the treatment of a vector order parameter~\cite{degennes1993}
\begin{equation}
\begin{split}
\label{freeE_pol}
\mathcal{F}^{pol}[\phi, \mathbf{P}] = \int d^{2}r\, \biggl\{ -&\frac{\alpha}{2} \frac{(\phi-\phi_{cr})}{\phi_{cr}}\left|\mathbf{P}\right|^2 +  \\ &  \frac{\alpha}{4}\left|\mathbf{P}\right|^{4}+\frac{k_P}{2}(\nabla\mathbf{P})^{2} \biggl\} .
\end{split}
\end{equation} 
Here the first two terms define the bulk properties of the liquid crystal that is confined in those regions where $\phi>\phi_{cr}$,  while the term proportional to $k_P$ describes the energy cost due to elastic deformation of the polar pattern.
Finally, the last  term on the right hand side of Eq.~\eqref{freeE} defines the anchoring interaction between the concentration and the polarization field
\begin{equation}
\label{freeE_anch}
\mathcal{F}^{anch}[\phi,\mathbf{P}] = \int d^{2}r\,\beta\mathbf{P}\cdot\nabla\phi .
\end{equation} 
This contribution favours homeotropic anchoring of $\mathbf{P}$ at the interfaces. In particular, when $\beta>0$, the polarization points towards the decreasing values of $\phi$. 
The ground state defined by the free energy functional $\mathcal{F}[\phi, \mathbf{P}]$ is characterized by a transition from the isotropic towards the lamellar phase for $a<k^2_\phi/4c+\beta^2/k_P$, with the polarization $\mathbf{P}$ confined in only one of the two phase and normally anchored to the interfaces.

The dynamic equations governing the physics of the system are
\begin{align}
&\rho\left(\frac{\partial}{\partial t}+\mathbf{v}\cdot\nabla\right)\mathbf{v}  =  -\nabla P+\nabla\cdot (\tilde{\sigma}^{pass}+\tilde{\sigma}^{act}),\label{eqn:navier_stokes}\\
&\frac{\partial \phi}{\partial t}+\nabla\cdot\left(\phi\mathbf{v}\right)=\nabla\left( M\nabla\frac{\delta \mathcal{F}}{\delta \phi}\right),\label{eqn:conv_diff}\\
&\frac{\partial\mathbf{P}}{\partial t}+\left(\mathbf{v}\cdot\nabla\right)\mathbf{P}=-\tilde{\Omega}\cdot\mathbf{P}+\xi\tilde{D}\cdot\mathbf{P}
-\frac{1}{\Gamma}\frac{\delta \mathcal{F}}{\delta\mathbf{P}},\label{eqn:beris}
\end{align}
in the limit of incompressible fluid (which is a good approximation for active fluids). Eq.~\eqref{eqn:navier_stokes} is the Navier-Stokes equation, where $P$ is the ideal pressure and the stress tensor has been divided in a passive and an active contribution. The former is due to dissipative (viscous) and reactive phenomena resulting from the dynamics of the two order parameters. Its explicit expression is given by the sum of three terms
\begin{equation}
\label{eqn:pass_stress_tensor}
\tilde{\sigma}^{pass} = \tilde{\sigma}^{visc} + \tilde{\sigma}^{bm} + \tilde{\sigma}^{pol},
\end{equation}
whose explicit definition is given in the Appendix.
The active contribution has phenomenological origin and it is the only term not stemming from the free energy.
It accounts for the local stress exerted by the active particles and is given by~\cite{ramaswamy2010,marchetti2013}
\begin{equation}
\label{eqn:active_stress}
\sigma_{\alpha\beta}^{act}=-\zeta \phi \left(P_{\alpha}P_{\beta}-\frac{1}{2}|{\bf P}|^2\delta_{\alpha\beta}\right),
\end{equation}
where $\zeta$ is the activity parameter, positive for extensile systems and negative for contractile ones.

By assuming the amount of active material to be conserved, $\phi$ evolves according to a convection-diffusion equation (Eq.~\eqref{eqn:conv_diff}) in which $M$ is the mobility and $\mu=\delta \mathcal{F}/\delta\phi$ is the chemical potential.

The evolution of the polarization field  is governed by the Ericksen-Leslie equation for a vector order parameter (Eq.~\eqref{eqn:beris}), where $\Gamma$ is the rotational viscosity, while $\tilde{D}$ and $\tilde{\Omega}$ respectively represent the symmetric and the antisymmetric part of the velocity gradient tensor $\partial_\alpha v_\beta$. 

The simulations were performed on a bi-dimensional lattice, initializing the concentration uniformly as $\phi=\phi_{cr}$.
The polarization was randomly initialized both in orientation and intensity.
We confined the system in a channel of width $L$ with no-slip boundary conditions at the top and the bottom walls ($y = 0$ and $y = L$). These conditions were implemented in the LB algorithm by bounce-back boundary conditions and periodic boundary conditions in the $y$ direction. The flow is driven by moving walls, with velocity $v_w$ for the top wall and $-v_w$ for the bottom wall. This gives a shear rate $\dot{\gamma}= 2v_w /L$.
Neutral wetting boundary conditions were enforced by requiring on the wall sites that:
\begin{equation}
\label{eqn:neutral_wetting}
\nabla_\perp \mu(y=0,L) = 0, \qquad \nabla_\perp \nabla^2 \phi(y=0,L)	 =0
\end{equation}
where $\nabla_\perp$ denotes the partial derivative computed normally to the walls and directed towards the bulk of the system. The first condition ensures density conservation, and the second determines the wetting to be neutral. 
Concerning the boundary conditions for the polarization we impose strong tangential boundary conditions
\begin{equation}
\label{eqn:P_boundary}
P_\perp(y=0,L)=0, \qquad \nabla_\perp P_\parallel(y=0,L)=0,
\end{equation}
where $P_\perp$ and $P_\parallel$ denote, respectively, the normal and tangential components of the polarization field with respect to the walls. This is because, according to experiments, both bacteria and cytoskeletal extracts tend to orient along the wall directions when they are close to the boundaries.

Other numerical details concerning the numerical method and the mapping to physical values are given in the Appendix.
 
\section*{Results}

\subsection*{Asymmetry and soft channel formation in contractile emulsions}

\begin{figure*}[t]
\centering
	\includegraphics[width = 1.0\textwidth]{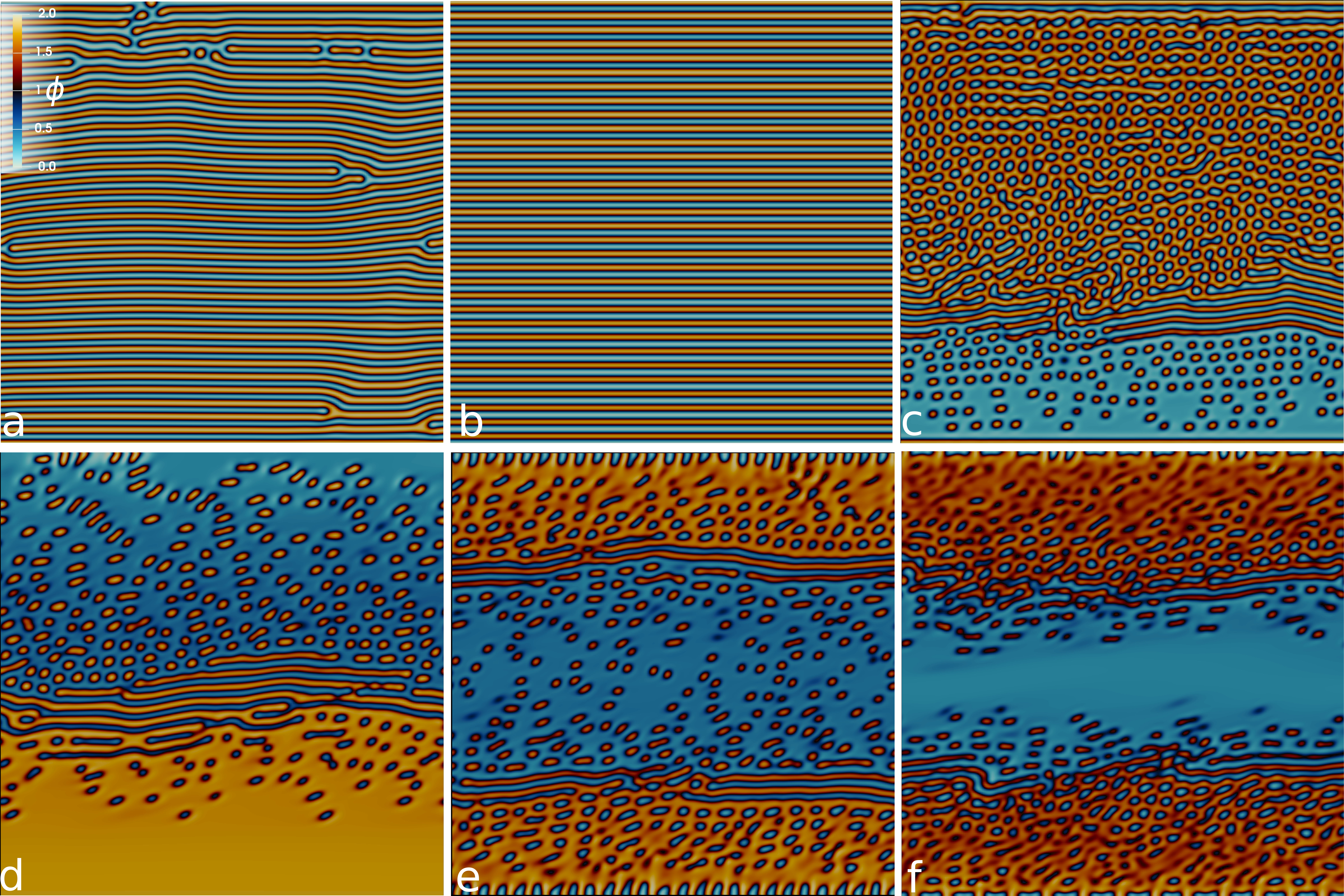}
\caption{\textbf{Contour plots of $\phi$ at varying activity and shear rate.} Top and bottom row correspond to the case at shear rate $\dot{\gamma}=0.78 \times 10^{-4}, 7.1 \times 10^{-4}$ respectively. Left panels correspond to the passive limit ($\zeta=0$), while middle and right columns show the configuration for two cases at intermediate ($\zeta=-0.001$) and strong ($\zeta=-0.005$) contractile activity. The color range of the contour plot, shown in panel~a, is the same for all figures and is defined as follows: red-yellow (active phase) if $\phi>1$,  dark blue-blue (passive phase) if $\phi<1$.}
\label{fig:fig.1}
\end{figure*}

In order to analyze the behavior of our system, we performed a systematic scanning of both contractile activity and shear rate. In this Section we will present  the morphological properties, summarized in Fig.~\ref{fig:fig.1}, showing some examples of the different regimes occurring in our system.
At equilibrium ($\zeta=0$) the free energy in Eq.~\eqref{freeE} prescribes the formation of a lamellar configuration for the concentration field. In absence of forcing (either external or internal), and starting from a mixed state, large domains of well aligned lamellae develop throughout the system. Complete ordering  is however unlikely to be reached due to persistent dislocations that diffusion or backflow are not able to flush away.
If a weak shear rate is externally imposed ($\dot{\gamma} \lesssim 0.8 \times 10^{-4}$), the lamellar texture aligns in the flow direction as shown in panel~(a) of Fig.~\ref{fig:fig.1}. In this case, dislocations in the lamellar pattern still persist in the long time dynamics. This changes when an active emulsion is considered. By keeping fixed the shear rate and switching on weak contractile activity ($|\zeta| \lesssim 4\times 10^{-3}$), a completely ordered configuration is achieved with no defects in the lamellar arrangement, regardless of the initial conditions (see Fig.~\ref{fig:fig.1}(b)). This is due to the mutual interaction between the external shear and the activity: while the former has the effect of aligning the pattern to the flow, the latter provides a non-vanishing energy contribution which keeps on stirring the fluid and eventually leads to the elimination of defects.

A further increment of the intensity of the active forcing engenders an important morphological transition. When activity is strong enough, the lamellar texture becomes unstable to deformations of the polar pattern. The underlying mechanism is the generic instability of active contractile materials to splay~\cite{marchetti2013,ramaswamy2010}. This causes the lamellae to disrupt with the consequent development of an active matrix populated by passive droplets, whose formation is energetically favoured as they allow for relieving elastic stresses~\cite{bonelli2019}. More interestingly, this phenomenology --which was already observed in unconfined systems-- is accompanied by a symmetry breaking in the gradient direction, as shown in panel~(c) of Fig.~\ref{fig:fig.1}. The active matrix gathers on one side of the channel, while a layer of isotropic fluid is formed on the other side, with some active features scattered  with no definite pattern. 
As we will discuss in more depth in the following Section, for low shear rate and intense activity, the symmetry breaking is the result of the advective transfer of active material from one side of the channel to the other.

To investigate the role of the external forcing in such dynamical process, we considered the case of stronger shear rates (see panels~(d)-(f) in Fig.~\ref{fig:fig.1}). We found that, even a passive emulsion can undergo a similar morphological transition to an asymmetric state, as suggested by the contour plot of the concentration field in panel (d). Nevertheless, the matrix of polar liquid crystal is now more uniform if compared to its active analogue, and droplets of isotropic fluid only develop far from the wall, where the polarization roughly aligns at the prescribed Leslie angle~\cite{markovich2019} ($\theta_L \simeq 0.21 \text{rad}$)\footnote{Under shear flow, a passive polar gel is tilted with respect to the flow at an angle $\theta_L=\sqrt{(\xi-1)/(\xi+1)}$, known as Leslie angle. We checked that, in the passive limit, the  polarization aligns at the Leslie angle when the symmetry is broken and a liquid crystal slab is formed at one of the walls.}. This suggests the dynamical process is led  by an effective elastic interaction which overtakes free energy relaxation, thus favouring the formation of large domains where the liquid crystal is uniformly aligned, more than preserving the lamellar texture. 
We considered the case of an active contractile emulsion under a strong shear rate ($\dot{\gamma}=7\times 10^{-4}$). Remarkably, under these conditions, activity prevents the development of any asymmetry (Fig.~\ref{fig:fig.1}(e)). Two active layers form at the walls, with the proliferation of passive droplets, while a channel of isotropic fluid, populated by small active domains, develops in the centre of the system. As activity is increased (see panel~(f)) the channel shrinks and gets rid of polar droplets, while the active layers widen, due to the absorption of dispersed active material floating in the isotropic fluid.
\begin{figure*}[t]
\centering
	\includegraphics[width = 1.0\textwidth]{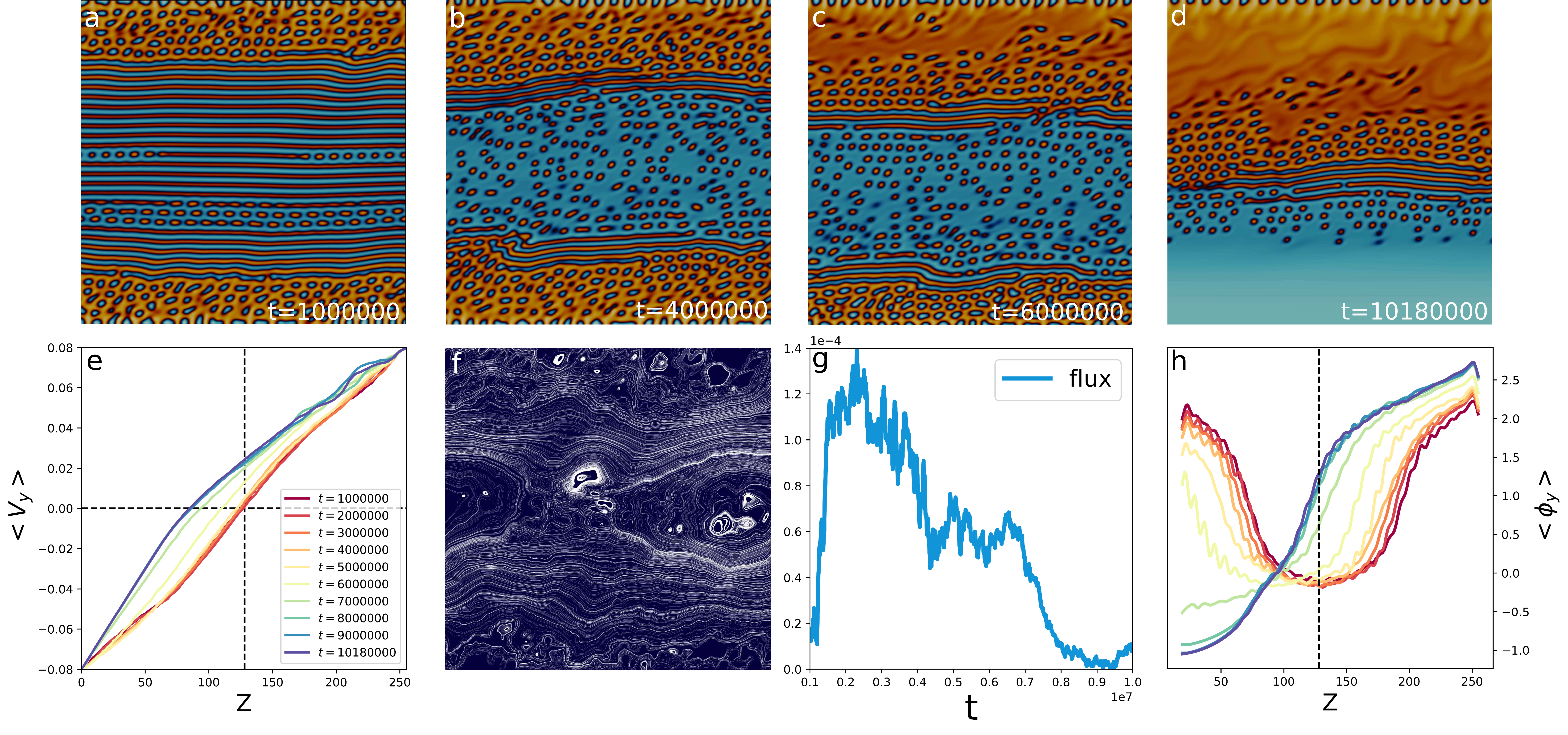}
\caption{\textbf{Activity induces symmetry breaking} Panels (a)-(f) show different stages of the development of the asymmetry for $\zeta=-0.002$ and $\dot\gamma=6.2\times 10^{-4}$, from the pearlification of the lamellar pattern (a), isotropic channel formation (b), shrinking of one of the two active layers (c) to the fully developed asymmetric configuration (d). Panel (e) shows the evolution of the $v_y$ profiles averaged in the channel direction $y$. Panel (f) shows the streamlines of the reduced velocity field $\mathbf{\tilde{v}}=\mathbf{v}-\mathbf{V}$, where $\mathbf{V}=(-v_w+\dot{\gamma}z)\mathbf{j}$ is the shear flow. Panel (g) shows the advective flux of the concentration field, computed  as the amount of active material that flows across the mid-line of the simulation box in a lapse of $10^4$ iterations ($\Phi= \int \text{d}t \text{d}y \phi v_z$). Panel (h) shows the temporal evolution of the concentration profile, from the lamellar (bordeaux), to the final asymmetric (blue) configuration.}
\label{fig:fig.2}
\end{figure*}
To summarize, the asymmetry may develop due to both activity and external shear separately, as the result of an effective elastic interaction which drives the growth of the liquid crystal domain sustained by the advection of the concentration field. Nevertheless, when both active and external forcing are strong enough, the competition of the two mechanisms leads to a symmetric configuration (see also the phase diagram in Fig.~\ref{fig:fig.5}(a)). In the following we will refer to the shear as \emph{weak} whichever the lamellar phase is preserved ($\dot{\gamma} \lesssim 0.8 \times 10^{-4}$); analogously \emph{strong} shear corresponds to a situation where the formation of the soft channel is favoured ($\dot{\gamma} \gtrsim 7 \times 10^{-4}$).

In the next sections we will first thoroughly address the dynamical mechanisms driving the instability in the case of an active emulsion at strong activity and shear rate. Later on we will discuss the formation of soft channels, then the stability of the asymmetric state.

\subsection*{The dynamics of symmetry breaking}

In the previous Section we commented on the development of symmetry breaking in the concentration pattern occurring for weak shear rates and strong enough contractile activity. Panels (a)-(d) of Fig.~\ref{fig:fig.2} show the different stages of the process for the case with $\zeta=-0.002$ and $\dot\gamma=6.2\times 10^{-4}$. Starting from a totally mixed configuration, the system rapidly relaxes into a lamellar pattern. 
Due to the imposed shear, lamellae grow increasingly thinner until they disrupt, into a pearlification texture (panel~(a)). Concurrently, the excess of active material expelled from the bulk is progressively captured by the thick active matrix adhering on the walls. Initially both layers progressively widen (panel~(b)), even if the absorption process may proceed at different speed, thus initiating the symmetry breaking. 

\begin{figure*}[t]
\centering
	\includegraphics[width = 0.95\textwidth]{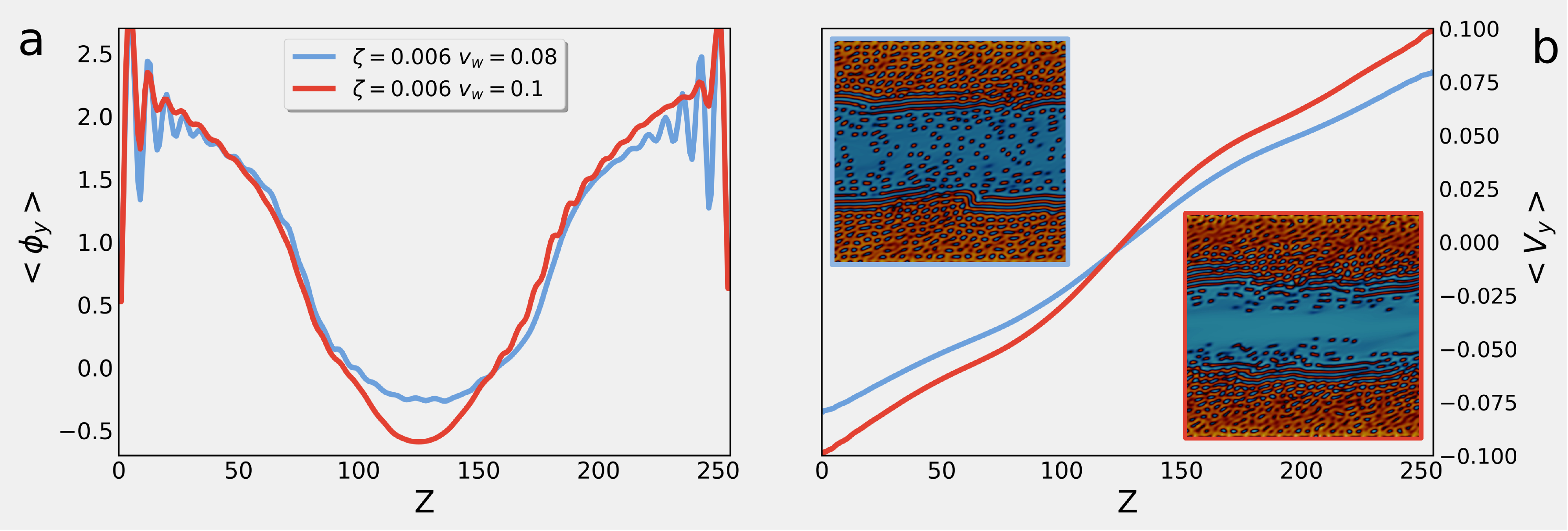}
\caption{\textbf{Soft channel formation} Panel (a): Concentration profiles averaged in the flow direction at fixed activity $\zeta=-0.006$, for two values of the wall velocity $v_w=0.08$  (blue line) and $0.1$ (red line), corresponding to $\dot\gamma=6.3\times 10^{-4}$, $7.8\times 10^{-4}$, respectively. Profiles oscillate in proximity of the walls and progressively decrease as the interface between the active layer and the isotropic channel is approached. At higher shear rate the profile is peaked, while it is smoother for less intense external forcing, showing that in the former case the soft channel has less suspended active droplets with respect to the latter. Contour plots for the two cases (respectively framed with a blue and a red margin) are shown as insets of panel (b) where the velocity profiles averaged in the channel direction are shown. Both of them exhibit two shear banding in correspondence of the isotropic/liquid crystal interface. These are more pronounced for higher shear rate, where the channel is emptier.}
\label{fig:fig.3}
\end{figure*}

When the lamellar texture disappears from the centre of the system, an emulsion of small active droplets remains suspended in the bulk of isotropic fluid. These are advected by the flow, whose structure is determined by both the imposed shear and the distribution of active material. To utterly analyse the flow structure, we plotted the reduced velocity field $\mathbf{\tilde{v}}$ in panel~(f) --namely the difference between the actual velocity $\mathbf{v}$ and the imposed shear flow $\mathbf{V}$-- which exhibits a large convection roll. This has the important effect of establishing a circulating flux of active material, so that more and more active droplets are released in the channel from the melting layer, advected by the flow and finally absorbed by the opposite layer that keeps on growing. 
This process is characterized by a net flux $\Phi$ of the concentration field~(see Fig.~\ref{fig:fig.2}(g)). The flux here is computed as the amount of active material that flows across the mid-line of the simulation box in a lapse of $10^4$ iterations ($\Phi= \int \text{d}t \text{d}y \phi v_z$).
We notice that in the early dynamics --corresponding to the stage at which the  lamellar pattern undergoes pearlification leading to the formation of the isotropic channel-- the concentration flux is approximately null, while it attains a positive value during the shrinking of the bottom layer, showing that convection is responsible for the transfer of active material from the lower to the upper half of the channel. Finally, the flux decreases until it drops back to $0$,  and the system settled in the final asymmetric configuration of panel~(d), characterized by a channel of isotropic fluid in the lower half and an active layer in the upper part of the system.

To quantitatively address the rheological features of our system, we measured the concentration profiles $\langle \phi_\parallel \rangle$ and the velocity profiles $ \langle v_\parallel \rangle $ averaged along the channel associated with the different stages of the symmetry breaking dynamics (see panels~(e) and~(h)). Two shear bandings are clearly visible as long as the system is symmetric and are associated with corresponding concentration banding (red and yellow lines). We observe that the effective shear rate is considerably lower in the regions where the inverse active emulsion is well developed, while it increases significantly in the centre of the channel. This behavior is related to the active nature of the matrices adhering to the walls. Indeed, the active liquid crystal produces in these regions a spontaneous flow which  adds to the externally imposed shear flow, eventually thwarting it.
As the bottom active layer melts, the velocity profile gets smoother and smoother so that the shear banding in proximity of the bottom wall progressively disappears and completely vanishes when the system becomes fully asymmetric.

The late time configuration is characterized by strong concentration and shear banding, visible in the blue profiles in panels~(h) and~(e). In particular, we observe that the local shear rate is greater in the isotropic domain, than in the active region, where the slope of the blue velocity profiles is lower --thus indicating that contractile activity induces an effective viscosity increase, \emph{i.e.} active thickening.

\subsection*{Soft channel formation and active thickening}

In the previous Section we discussed how activity triggers the breaking of the top/bottom symmetry, due to the development of convection rolls --sustained by the energy injected by the active constituents-- that advect the concentration field, leading to a net flux that moves material from one side of the channel to the other.
This transfer mechanism is effective for strong active forcings at low shear rates, since the strength of the activity-induced flows is comparable with that of the external forcing. But what happens if we are to raise the shear rate?

In this case, the relative significance of the active flows with respect to the externally imposed shear is highly suppressed and the convection rolls --responsible for the advection of the concentration-- progressively disappear, thus leading to a situation where symmetry breaking would not occur.

To show the behavior of our system, we now consider the case where both shear and active forcing are strong. The system, initialized in the mixed configuration, undergoes a similar early-time dynamics as described in the previous Section: first a lamellar pattern is formed, then the texture is progressively destroyed due to pearlification instabilities. Two layers, characterized by a suspensions of passive droplets in an active background, develop at the walls and progressively thicken due to absorption of the active material, resulting in a symmetric configuration (see contour plots in the inset of Fig.~\ref{fig:fig.3}(b)), where a \emph{soft} channel of isotropic fluid forms in the centre of the system with a few suspended active features.
This configuration is found to be stable even at long simulation times, since the strong shear flow prevents convection to set up an effective concentration flux.
Interestingly, the shear rate plays an important role in controlling the amount of active material dispersed in the isotropic channel. Indeed, as the external forcing is increased the number of small active droplets progressively lowers, in a similar fashion to what happens when contractile activity is increased (see panels (e) and (f) of Fig.~\ref{fig:fig.1}). More quantitatively, we plotted in Fig.~\ref{fig:fig.3}(a) the concentration profiles for two values of $v_w$, by keeping fixed the activity ($\zeta=-0.006$). 
We observe that, in both cases, the $\phi$ profiles exhibit strong fluctuations at the boundaries, due to the dynamical nucleation of the passive droplets via a mechanism resembling Ostwald ripening. In this region, active material is denser for the case at higher shear rate. As the interface between the active layer and the isotropic channel is approached, the profiles drastically decrease to attain a minimum in the centre of the channel. Nevertheless, while in the case at $v_w=0.08$ (blue curve) the profile flattens in the isotropic region, due to the presence of active droplets in the background, for stronger external forcing ($v_w=0.1$, red curve) the profile is much more peaked. 
These  morphological features have an important effect on the velocity profiles, shown in panel (b) of Fig.~\ref{fig:fig.3}. Analogously to what happens in the case of asymmetric configurations discussed in the previous Section, shear rate is more pronounced in the isotropic region, than in the active layers, where the effective viscosity is higher. However, this local thickening effect is also affecting the global rheological properties of the system, leading to the increase of the total effective viscosity with the intensity of the active forcing (see Fig.~\ref{fig:fig.5}(b)).
\begin{figure*}[t]
\centering
	\includegraphics[width = 0.95\textwidth]{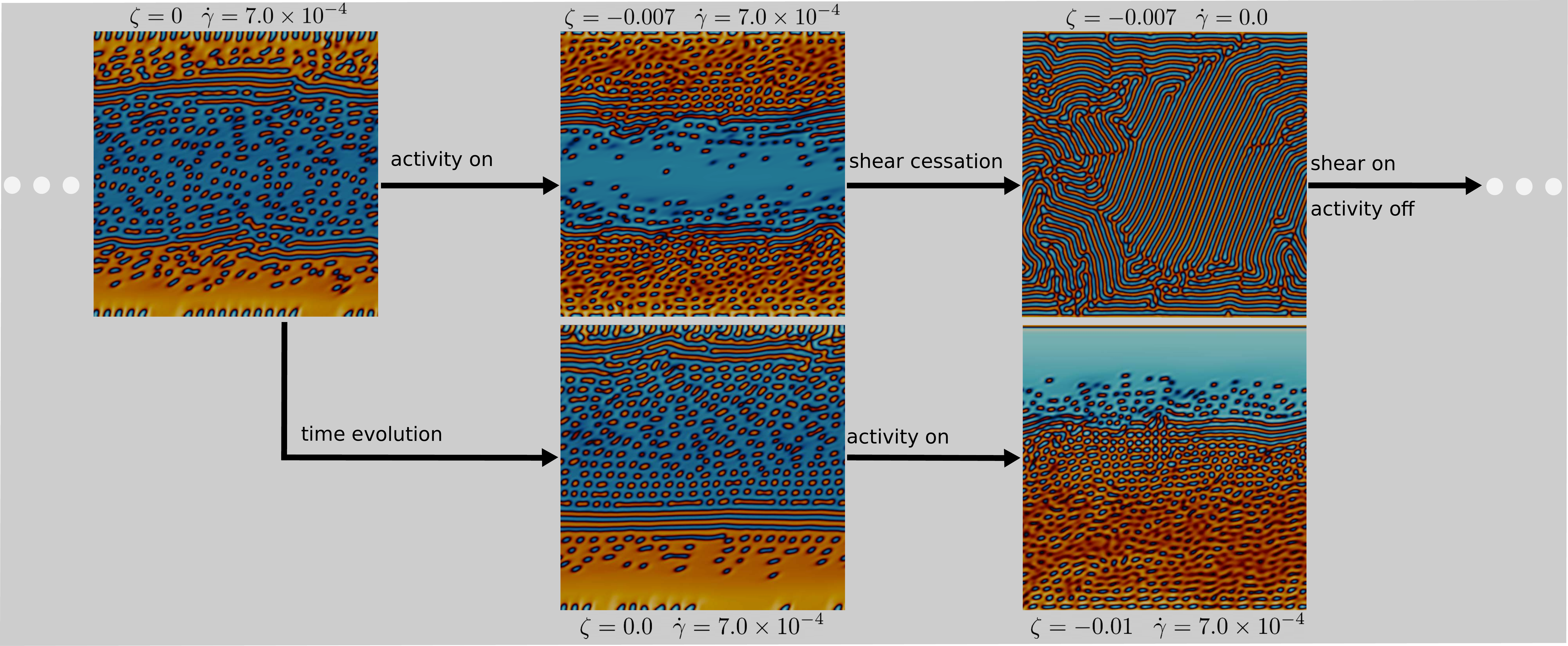}
\caption{\textbf{Soft channel stability} Top row: starting from an mixed state the system is evolved in absence of activity with shear ratem 	$ \dot\gamma=7.0\times 10^{-4}$ until an appreciable asymmetry is developed (top left panel). Turning on contractile activity allows to stabilize an isotropic channel in the centre of the system (top centre panel). This configuration evolves towards a lamellar pattern if the shear is turned off (top right panel). The initial configuration can be restored by switching of the activity and turning the shear on again. Bottom row: if the asymmetry is fully developed when the activity is turned on (bottom centre panel) activity is not able to stabilize the isotropic channel and the final configuration is asymmetric (bottom right panel). The lamellar pattern can be restored from this state turning switching the activity off.}
\label{fig:fig.4}
\end{figure*}
This concludes the characterization of the dynamical regimes. A final question needs to be answered: does the system exhibit story dependence? Are there any hysteresis features? 
We will address this subject in the following Section and we will show how the channel configuration is effectively stabilized by intense contractile activity even in presence of a partial asymmetry, given that this is not far developed.

\subsection*{Hysteresis features of soft channel formation}

In order to investigate the stability of the symmetric configurations we performed a dynamical study varying activity and shear rate over time.
We first initialized the system in the mixed state and we started the simulation imposing a shear rate ($\dot\gamma=7.0\times 10^{-4}$) such to obtain an asymmetric configuration. As the system develops an appreciable asymmetry (as shown in the top left panel of Fig.~\ref{fig:fig.4}) we turned on the activity ($\zeta=-0.007$). We chose a value corresponding to a case characterized by the formation of an isotropic channel and we waited for the system to settle into a stable configuration. Importantly, the final configuration is symmetric (central top panel), proving that strong contractile activity is able to stabilize the channel, even starting from an asymmetric state. Importantly, this configuration is sustainable only if an external shear flow is imposed. Indeed, switching off the external forcing leads to the lamellar state (top right panel of Fig.~\ref{fig:fig.4}) that one would obtain if the system were initialized in the mixed state. The starting asymmetric configuration can be now obtained by switching the activity off and the shear rate on at its original value. This dynamical test shows that contractile activity is fundamental to stabilize the isotropic channel, even if an applied external forcing is necessary for this configuration to be stable. Nevertheless, our system still exhibit some history dependence. Indeed, in the first stage of our test, we turned on the activity when the asymmetry was not fully developed. What would happen if we perform the same test starting from a more asymmetric state, as the one shown in the bottom central panel of Fig.~\ref{fig:fig.4}? In this case the amount of active material in the upper part of the system is not enough to set up an effecting counteracting flow and the system develops an asymmetric configuration, analogous to those observed at lower shear rate (bottom right panel). Nevertheless, by switching off the shear rate once more, it is possible to recover the lamellar state.

\section*{Discussion and Conclusions}
In this work we analysed the morphology and rheology of a mixture of an isotropic fluid and a polar active fluid under shear.  We implicitly modelled the action of a surfactant which favours emulsification in the absence of shear. By varying the strength of the activity parameter (which measures the force generated internally by the system) and the imposed shear rate, we showed that the system can attain three main possible morphologies (see phase diagram in Fig. 5(a)). For low shear rates ($\dot{\gamma} \lesssim 0.8 \times 10^{-4}$), the emulsion settles into a lamellar pattern. Defects in the lamellar ordering in this regime are eliminated by increasing activity, giving rise eventually, for sufficiently strong activity, to a state where lamellae align along the direction of the channel. For larger shear rates, we observe a morphological transition to a completely different state, characterised by phase separation where an active matrix pierced by a passive droplet emulsion coexists with an isotropic fluid with active droplets. This phase separation is driven by a convective flux which relocates active material from one side of the system to the other; this regime is additionally characterised by a strong concentration and shear banding as the two phases flow differently and have quite distinct average composition. There are two morphologies corresponding to the phase separated state. At low activity, the structure is asymmetric, and the isotropic phase and the active matrix migrate to different sides. For large activity, the structure is remodelled to create an internal isotropic channel flanked by two soft active matrices wetting the walls. Our simulations suggest that the formation of a symmetric channel requires activity, to dissolve the active droplets in the middle of the isotropic phase.
\begin{figure*}[t]
\centering
		\includegraphics[width = 0.95\textwidth]{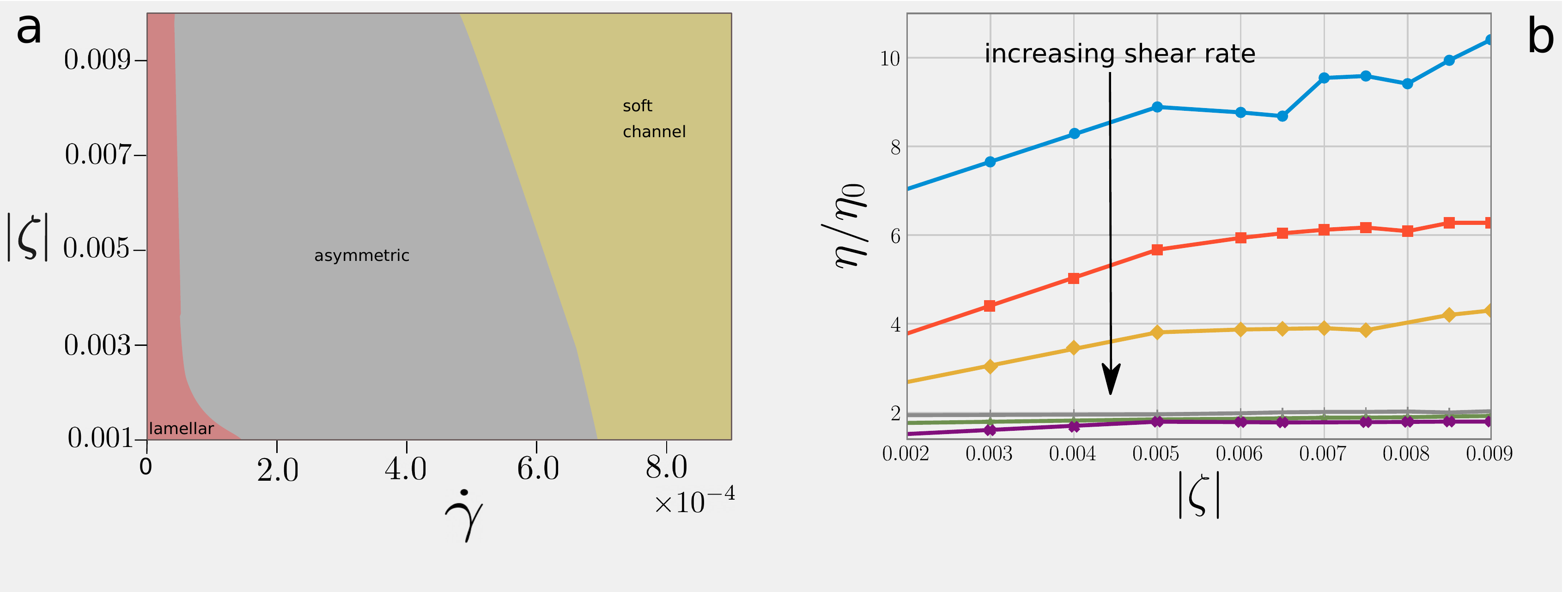}
\caption{\textbf{Morphological regimes and rheological properties}. Panel (a): Phase diagram in the plane generated by the two control parameters (the strength of the active energy injection and shear rate). Panel (b) shows the behavior of the measured viscosity $\eta$ (computed as $\langle \sigma^{tot}_{yz} \rangle /\dot{\gamma}$) normalized with respect the nominal viscosity of the fluid at varying the intensity of contractile activity $\zeta$, for different values of shear rate $\dot\gamma$. The corresponding values of shear rate from top (blue curve) to  bottom (violet curve) are: $\dot\gamma=1.95\times 10^{-5},7.81\times 10^{-5},1.56\times 10^{-4}, 4.68\times 10^{-4}, 6.25\times 10^{-4},7.81\times 10^{-4}$.  }
\label{fig:fig.5}
\end{figure*}
To characterise the rheological properties of the system we plotted in Fig. 5 the behaviour of the apparent viscosity $\eta$, normalised by the background fluid viscosity (i.e., the one entering the purely viscous part of the stress tensor). At constant activity, the viscosity lowers when shear rate is increased, so the system is shear thinning. Additionally, we observe that activity leads to an increase in viscosity, as in single-phase contractile fluids~\cite{marenduzzo2007}, an effect which we refer to as active thickening. Active thickening is more visible for small shear rates, while it becomes negligible when the wall velocity is very strong, as the externally imposed flow dwarfs the active one.

Our results show that activity and an imposed flow can be harnessed to enhance the self-assembly potential of active systems, and specifically to create soft microfluidic channels by simply mixing suitable ingredients. It would be of interest to perform fully 3D simulations of this system to see how spatial dimensionality affects pattern formation, and whether a soft channel can be self-assembled in this case as well. Additionally, it would be desirable to study active self-assembly in active mixtures and emulsions experimentally. A direct replication of the ideas proposed here may be challenging, as it would require the creation of a two-dimensional emulsion. However, similar studies may be possible with actomyosin-water emulsions in the presence of a surfactant. Our results suggests that systems like these are likely to display a very rich and intriguing phenomenology.

\section*{Numerical details and mapping to physical values}
The stress tensor appearing at the right-hand side of the Navier-Stokes equation has been divided in two parts, respectively addressed as \emph{passive} $\tilde{\mathbf{\sigma}}^{pass}$ and \emph{active} $\tilde{\mathbf{\sigma}}^{act}$. The first is in turn the sum of dissipative and reactive contributions. 
The viscous term is given by
\begin{equation}
\label{eqn:viscous_stress}
\sigma_{\alpha\beta}^{visc}=\eta(\partial_{\alpha}v_{\beta}+\partial_{\beta}v_{\alpha}),
\end{equation}
where $\eta$ is the shear viscosity. The second term in Eq.~\eqref{eqn:pass_stress_tensor} is an interfacial contribution and is given by
\begin{equation}
\label{eqn:interface_stress}
\sigma_{\alpha\beta}^{bm}=\left( f-\phi\frac{\delta \mathcal{F}}{\delta\phi} \right)\delta_{\alpha\beta} - \frac{\partial f}{\partial( \partial_\beta\phi)} \partial_\alpha \phi,
\end{equation}
where $f$ is the free energy density.
The last term term gives the stress due to deformations of the liquid crystal pattern. Its explicit form is
\begin{equation}
\label{eqn:elastic_stress}
\sigma_{\alpha\beta}^{pol}=\frac{1}{2}(P_{\alpha}h_{\beta}-P_{\beta}h_{\alpha})-\frac{\xi}{2}(P_{\alpha}h_{\beta}+P_{\beta}h_{\alpha})
-\kappa\partial_{\alpha}P_{\gamma}\partial_{\beta}P_{\gamma},
\end{equation}
where $\textbf{h}=\delta \mathcal{F}/\delta\textbf{P}$ is the molecular field~\cite{degennes1993} and the shape factor $\xi$ --the shap factor-- controls the anisotropy of the liquid crystal molecules and its response to a shear flow. In particular, it selects rod-like particles if positive or disk-like ones if negative. Moreover, the polarization field exhibits flow aligning or flow thumbling features under shear if $|\xi|>1$ or $|\xi|<1$, respectively.

Eqs.~\eqref{eqn:navier_stokes}-\eqref{eqn:beris} are numerically solved by means of a well validated hybrid lattice Boltzmann method (in the limit of incompressible flow). The
Navier-Stokes equation was solved through a predictor-corrector LB scheme, while the evolution equations for the order parameters $\phi$ and $\mathbf{P}$ were integrated through a predictor-corrector finite-difference algorithm
implementing first-order upwind scheme and fourth order accurate stencils for space derivatives.
Simulations are performed on a two-dimensional square lattice (D2Q9) whose linear size ranges from $L=256$ to $L=512$. The system is initialized in a mixed state, with $\phi$ uniformly distributed between $1.1$ and $0.9$. The concentration $\phi$ ranges from $\phi=0$ (passive phase) to $\phi\simeq 2$ (active phase), that correspond to  the two minima of the double-well potential. The starting polarization field $\textbf{P}$ is randomly distributed between 0 (passive phase) and 1 (active phase). 
Unless otherwise stated, parameter values are $a=4\times 10^{-3}$, $k_\phi=-6\times 10^{-3}$, $c=10^{-2}$, $\alpha=10^{-3}$, $k_P=10^{-2}$, $\Gamma=1$, $\xi=1.1$, $\phi_0=2$, $\beta=10^{-2}$, $\eta=1.67$.
By following previous studies~\cite{elsen}, an approximate relation between simulation units and physical ones (such as those of a contractile active gel) can be obtained by using as length-scale, time-scale and force-scale respectively the values $L=1\mu$m, $\tau=10$ms and $F=1000$nN (see Table.~\ref{table1}). 
Throughout our simulations the Reynolds number, for the case in which the droplets are observed, is evaluated in terms of the average droplet radius, of the viscosity and of the velocity of the fluid. It remains below $0.25$, a value in which inertial effects are indeed negligible.

\begin{table}[htbp]
\centering
\caption{Typical values of the physical quantities used in the simulations.}
\label{table1}
\vskip 0.3cm

\begin{tabular}{p{3.8cm}|c|c}
Model variables and parameters & Simulation units & Physical units  \\
\hline
Effective shear viscosity, $\eta$                        & $5/3$                 & $1.67\, \mathrm{kPa}\mathrm{s}$ \\
Effective elastic constant, $\kappa$                     & $0.006$               & $6\, \mathrm{nN}$ \\
Shape factor, $\xi$                                      & $1.1$                 & dimensionless \\
Effective diffusion constant, $D=Ma$                     & $0.0004$              & $0.004\,\mu \mathrm{m}^{2}\mathrm{s}^{-1}$ \\ 
Rotational viscosity, $\Gamma$                           & $1$                   & $10\, \mathrm{kPa}\mathrm{s}$ \\
Activity, $\zeta$                                        & $0-0.01$              & $(0-100)\, \mathrm{kPa}$ \\
\end{tabular}
\end{table}

\section*{Acknowledgements}

Simulations have been performed at Bari ReCaS e-Infrastructure funded by MIUR through the program PON Research and Competitiveness 2007-2013 Call 254 Action I and at ARCHER UK National Supercomputing Service (http://www.archer.ac.uk) through the program HPC-Europa3. We thank E. Orlandini for very useful discussions. A.T. acknowledges funding from the European Research Council under
the European Union’s Horizon 2020 Framework Programme (No. FP/2014-2020) ERC Grant
Agreement No. 739964 (COPMAT).

\section*{Author contributions statement}

L.N. C., A. L., G. G., D. M., G.N., and A. T. designed and performed the research and wrote the manuscript.

\section*{Additional information}

\textbf{Competing financial interests:} The authors declare no competing interests. 

\bibliographystyle{abbrv}

\bibliography{refs} 

\begin{thebibliography}{10}

\bibitem{bonelli2019}
F.~Bonelli, L.~Carenza, G.~Gonnella, D.~Marenduzzo, E.~Orlandini, and
  A.~Tiribocchi.
\newblock Lamellar ordering, droplet formation and phase inversion in exotic
  active emulsions.
\newblock {\em Sci. Rep.}, 9:2801, 2019.

\bibitem{braz}
S.~{Brazovski{\v i}}.
\newblock Phase transition of an isotropic system to a nonuniform state.
\newblock {\em J. Exp. Theor. Phys.}, 41:85, 1975.

\bibitem{carenza19turb}
L.~Carenza, L.~Biferale, and G.~Gonnella.
\newblock Multiscale control of active emulsion dynamics.
\newblock {\em Phys. Rev. Fluids}, 5:011302, Jan 2020.

\bibitem{reviewepje}
L.~Carenza, G.~Gonnella, A.~Lamura, G.~Negro, and A.~Tiribocchi.
\newblock Lattice boltzmann methods and active fluids.
\newblock {\em Eur. Phys. J. E}, 42:81, 2019.

\bibitem{Carenza22065}
L.~N. Carenza, G.~Gonnella, D.~Marenduzzo, and G.~Negro.
\newblock Rotation and propulsion in 3d active chiral droplets.
\newblock {\em Proceedings of the National Academy of Sciences},
  116(44):22065--22070, 2019.

\bibitem{cates2015}
M.~Cates and J.~Tailleur.
\newblock {Motility-Induced Phase Separation}.
\newblock {\em Annu. Rev. Condens. Matter Phys.}, 6:219--244, 2015.

\bibitem{degennes1993}
P.~de~Gennes and J.~Prost.
\newblock {\em The physics of liquid crystals}.
\newblock The International series of monographs on physics. Oxford University
  Press, 2nd edition, 1993.

\bibitem{denniston2001}
C.~Denniston, E.~Orlandini, and J.~Yeomans.
\newblock Lattice boltzmann simulations of liquid crystal hydrodynamics.
\newblock {\em Phys. Rev. E}, 63:056702, 2001.

\bibitem{digregorio2018}
P.~Digregorio, D.~Levis, A.~Suma, L.~Cugliandolo, G.~Gonnella, and
  I.~Pagonabarraga.
\newblock {Full Phase Diagram of Active Brownian Disks: From Melting to
  Motility-Induced Phase Separation}.
\newblock {\em Phys. Rev. Lett.}, 121:098003, 2018.

\bibitem{dombrowski2004}
C.~Dombrowski, L.~Cisneros, S.~Chatkaew, R.~Goldstein, and J.~Kessler.
\newblock {Self-Concentration and Large-Scale Coherence in Bacterial Dynamics}.
\newblock {\em Phys. Rev. Lett.}, 93:098103, 2004.

\bibitem{doostmohammadi2018}
A.~Doostmohammadi, J.~Ign{\'e}s-Mullol, J.~Yeomans, and F.~Sagu{\'e}s.
\newblock Active nematics.
\newblock {\em Nat. Commun.}, 9, 2018.

\bibitem{doostmohammadi2019}
A.~Doostmohammadi and J.~Yeomans.
\newblock Coherent motion of dense active matter.
\newblock {\em The European Physical Journal Special Topics}, 227:2401, 2019.

\bibitem{PhysRevX.5.031003}
L.~Giomi.
\newblock Geometry and topology of turbulence in active nematics.
\newblock {\em Phys. Rev. X}, 5:031003, 2015.

\bibitem{Giomi20130365}
L.~Giomi, M.~Bowick, P.~Mishra, R.~Sknepnek, and M.~Marchetti.
\newblock Defect dynamics in active nematics.
\newblock {\em Philosophical Transactions of the Royal Society of London A:
  Mathematical, Physical and Engineering Sciences}, 372(2029), 2014.

\bibitem{PhysRevLett.101.198101}
L.~Giomi, M.~C. Marchetti, and T.~B. Liverpool.
\newblock Complex spontaneous flows and concentration banding in active polar
  films.
\newblock {\em Phys. Rev. Lett.}, 101:198101, Nov 2008.

\bibitem{gonnella2015}
G.~Gonnella, D.~Marenduzzo, A.~Suma, and A.~Tiribocchi.
\newblock Motility-induced phase separation and coarsening in active matter.
\newblock {\em C. R. Phys.}, 16:316, 2015.

\bibitem{gonnella1998}
G.~Gonnella, E.~Orlandini, and J.~Yeomans.
\newblock {Lattice Boltzmann simulations of lamellar and droplet phases}.
\newblock {\em Phys. Rev. E}, 58:480--485, 1998.

\bibitem{guillamat2016}
P.~Guillamat, J.~Ign{\'e}s-Mullol, and F.~Sagu{\'e}s.
\newblock Control of active liquid crystals with a magnetic field.
\newblock {\em Proc. Natl. Acad. Sci. USA}, 113(20):5498, 2016.

\bibitem{guo2018}
S.~Guo, D.~Samanta, Y.~Peng, X.~Xu, and X.~Cheng.
\newblock Symmetric shear banding and swarming vortices in bacterial
  superfluids.
\newblock {\em Proc. Natl. Acad. Sci. USA}, 2018.

\bibitem{joanny2009}
J.-F. Joanny and J.~Prost.
\newblock {Active gels as a description of the actin-myosin cytoskeleton}.
\newblock {\em HFSP j.}, 3(2):94--104, 2009.

\bibitem{kruse2004}
K.~Kruse, J.-F. Joanny, F.~J\"ulicher, J.~Prost, and K.~Sekimoto.
\newblock {Asters, Vortices, and Rotating Spirals in Active Gels of Polar
  Filaments}.
\newblock {\em Phys. Rev. Lett.}, 92:078101, 2004.

\bibitem{DiLeonardo2010}
R.~D. Leonardo, L.~Angelani, D.~Dell{\textquoteright}Arciprete, G.~Ruocco,
  V.~Iebba, S.~Schippa, M.~Conte, F.~Mecarini, F.~D. Angelis, and E.~D.
  Fabrizio.
\newblock Bacterial ratchet motors.
\newblock {\em Proc. Natl. Acad. Sci. USA}, 107(21):9541--9545, 2010.

\bibitem{PhysRevLett.122.214503}
M.~Linkmann, G.~Boffetta, M.~C. Marchetti, and B.~Eckhardt.
\newblock Phase transition to large scale coherent structures in
  two-dimensional active matter turbulence.
\newblock {\em Phys. Rev. Lett.}, 122:214503, May 2019.

\bibitem{lopez2015}
H.~L\'opez, J.~Gachelin, C.~Douarche, H.~Auradou, and E.~Cl\'ement.
\newblock {Turning Bacteria Suspensions into Superfluids}.
\newblock {\em Phys. Rev. Lett.}, 115:028301, 2015.

\bibitem{marchetti2013}
M.~Marchetti, J.-F. Joanny, S.~Ramaswamy, T.~Liverpool, J.~Prost, M.~Rao, and
  R.~Simha.
\newblock Hydrodynamics of soft active matter.
\newblock {\em Rev. Mod. Phys.}, 85:1143, 2013.

\bibitem{marenduzzo2007}
D.~Marenduzzo, E.~Orlandini, and J.~Yeomans.
\newblock {Hydrodynamics and Rheology of Active Liquid Crystals: A Numerical
  Investigation}.
\newblock {\em Phys. Rev. Lett.}, 98:118102, 2007.

\bibitem{markovich2019}
T.~Markovich, E.~Tjhung, and M.~Cates.
\newblock {Shear-Induced First-Order Transition in Polar Liquid Crystals}.
\newblock {\em Phys. Rev. Lett.}, 122:088004, Feb 2019.

\bibitem{needleman2017}
D.~Needleman and Z.~Dogic.
\newblock Active matter at the interface between materials science and cell
  biology.
\newblock {\em Nat. Rev. Mater.}, 2, 2017.

\bibitem{negro2018}
G.~Negro, L.~Carenza, P.~Digregorio, G.~Gonnella, and A.~Lamura.
\newblock Morphology and flow patterns in highly asymmetric active emulsions.
\newblock {\em Physica A}, 503:464 -- 475, 2018.

\bibitem{prost2015}
J.~Prost, F.~J{\"u}licher, and J.-F. Joanny.
\newblock Active gel physics.
\newblock {\em Nat. Phys.}, 11(2):111, 2015.

\bibitem{ramaswamy2010}
S.~Ramaswamy.
\newblock {The Mechanics and Statistics of Active Matter}.
\newblock {\em Annu. Rev. Condens. Matter Phys.}, 1:323, 2010.

\bibitem{sanchez2012}
T.~Sanchez, D.~Chen, S.~Decamp, M.~Heymann, and Z.~Dogic.
\newblock Spontaneous motion in hierarchically assembled active matter.
\newblock {\em Nature}, 491:431--434, 2012.

\bibitem{simha2002}
R.~Simha and S.~Ramaswamy.
\newblock {Hydrodynamic Fluctuations and Instabilities in Ordered Suspensions
  of Self-Propelled Particles}.
\newblock {\em Phys. Rev. Lett.}, 89:058101, 2002.

\bibitem{Sokolov2009}
A.~Sokolov, M.~Apodaca, B.~Grzybowski, and S.~Aranson.
\newblock Swimming bacteria power microscopic gears.
\newblock {\em Proc. Natl. Acad. Sci. USA}, 107(3):969--974, 2010.

\bibitem{surrey2001}
T.~Surrey, F.~N{\'e}d{\'e}lec, S.~Leibler, and E.~Karsenti.
\newblock {Physical Properties Determining Self-Organization of Motors and
  Microtubules}.
\newblock {\em Science}, 292:1167, 2001.

\bibitem{thampi2016active}
S.~Thampi and J.~Yeomans.
\newblock Active turbulence in active nematics.
\newblock {\em Eur. Phys. J. Spec. Top.}, 225(4):651--662, 2016.

\bibitem{elsen}
E.~Tjhung, M.~Cates, and D.~Marenduzzo.
\newblock Nonequilibrium steady states in polar active fluids.
\newblock {\em Soft Matter}, 7:7453--7464, 2011.

\bibitem{tjhung2012}
E.~Tjhung, D.~Marenduzzo, and M.~Cates.
\newblock Spontaneous symmetry breaking in active droplets provides a generic
  route to motility.
\newblock {\em Proc. Natl. Acad. Sci. U.S.A.}, 109(31):12381--12386, 2012.

\bibitem{tjhung2015}
E.~Tjhung, A.~Tiribocchi, D.~Marenduzzo, and M.~Cates.
\newblock A minimal physical model captures the shapes of crawling cells.
\newblock {\em Nat. Commun.}, 6:5420, 2015.

\bibitem{wensink2012}
H.~Wensink, J.~Dunkel, S.~Heidenreich, K.~Drescher, R.~Goldstein, H.~Lowen, and
  J.~Yeomans.
\newblock Meso-scale turbulence in living fluids.
\newblock {\em Proc. Natl. Acad. Sci. USA}, 109, 2012.

\bibitem{Yeomans2014}
J.~Yeomans.
\newblock Playful topology.
\newblock {\em Nat. Mater.}, 13, 2014.

\bibitem{zhang2010}
H.~Zhang, A.~Be{\textquoteright}er, E.~Florin, and H.~Swinney.
\newblock Collective motion and density fluctuations in bacterial colonies.
\newblock {\em Proc. Natl. Acad. Sci. USA}, 107(31):13626--13630, 2010.

\end{thebibliography}

\end{document}